\documentclass[doublecol]{epl2}

\usepackage{amsmath}
\usepackage{units}
\usepackage{psfrag}
\usepackage{graphicx}

\newcommand{\e}{\mathrm{e}}

\newcommand{\rmd}{\mathrm{d}}

\newcommand{\Beta}{\mathrm{B}}

\begin{document}

\title{Spreading in Disordered Lattices with Different Nonlinearities}
\author{Mario Mulansky and Arkady Pikovsky} 
\institute{Department of Physics and Astronomy, Potsdam University, 14476 Potsdam, Germany, EU
}

\date{\today}

\abstract{We study the spreading of initially localized states in a nonlinear disordered lattice described by the nonlinear Schr\"odinger equation with random on-site potentials - a nonlinear generalization of the Anderson model of localization. We use a nonlinear diffusion equation to describe
the subdiffusive spreading. To confirm the self-similar nature of the evolution
we characterize the peak structure of the spreading states with help of R\'enyi entropies and in particular with the structural entropy. The latter is shown to remain constant over a wide range of time. Furthermore, we report on the dependence of the spreading exponents on the nonlinearity index in the generalized nonlinear Schr\"odinger disordered lattice, and show that these quantities are in accordance with previous theoretical estimates, based on assumptions of weak and very weak chaoticity of the dynamics. 
}


\maketitle
In disordered 1D lattices, all eigenmodes are exponentially localized due to Anderson localization \cite{Anderson-58}.
These models first appeared in the area of disordered electronic systems \cite{Kramer-MacKinnon-93,Lee-Ramakrishnan},
but they are also applicable to a wide variety of phenomena in a general context of waves (optical, acoustical, etc.) in disordered media~\cite{Sheng:2006,John:1984,Nature.390.671}. Localization effectively prevents spreading of energy in such situations.

By considering waves of large amplitudes, one faces \textit{nonlinearity} and  naturally encounters the question whether the nonlinearity destroys the localization or not.
Although this question has already been addressed numerically \cite{molina:1998,Kopidakis-Aubri-00,Pikovsky-Shepelyansky-08,flach:024101,Mulansky-Diplom,alpha_b_p:2009}, 
experimentally in BECs \cite{lye:070401,clement:170409,schulte:170411} and optical waveguides \cite{Nature.446.52,Lahini} as well as mathematically \cite{Wang-09},
 a full understanding is still elusive. It is easier to understand how nonlinearity destroys localization leading to thermalization~\cite{mulansky:2009} and self-transparency~\cite{Tietsche-Pikovsky} in short random lattices, than to analyze asymptotic regimes at large times in long lattices. For the latter setups a similarity between the quantum kicked rotor and a 1D Anderson model \cite{fgp,PhysRevLett.70.1787, Chirikov:1988} has provided an alternative realization of the effects of nonlinearity.

In this paper, we study structural properties of the spreading field in nonlinear disordered lattices, focusing on their dependence on the nonlinearity index.
Indeed, initial studies of the spreading of perturbations \cite{molina:1998,Pikovsky-Shepelyansky-08,flach:024101}
have been almost exclusively restricted to the behavior of the second moment of the distribution and of the participation number. These quantities, however, do 
not allow one to distinguish between all possible scenaria. In particular, the second moment of the distribution can grow due to a uniform spreading of the field, but also when localized packages move in opposite directions. Additionally, both these processes may coexist with some bursts that do not spread at all. In order to resolve these structural features in a statistical way, we apply for the first time a characterization of the spreading fields in nonlinear lattices with generalized R\'enyi entropies. For guidance, we compare the spreading properties with that of the nonlinear diffusion equation and study the relation between the effective diffusion index with the nonlinearity index of the original model.

Our basic model is described by the following generalization of the Discrete Anderson Nonlinear Schr\"odinger Equation (gDANSE):
\begin{equation} \label{eqn:gdanse}
   i\frac{\rmd}{\rmd t}\psi_n = V_n \psi_n + \psi_{n-1} + \psi_{n+1} + \beta |\psi_n|^{2\alpha} \psi_n\;.
\end{equation} 
Here $n=1,\dots,N$ is the lattice site index and $V_n$ is the uncorrelated random potential, chosen uniformly from the intervall $[-W/2, W/2]$.
The coefficient $\beta$  is proportional 
to the nonlinear strength (hereafter we assume a normalization $\sum_n|\psi_n|^2=1$). In this work, we consider only the case $\beta = 1$ and $W=4$.
The parameter $\alpha$, which we call nonlinearity index, is a novelity compared to the standard DANSE model with $\alpha = 1$  \cite{Pikovsky-Shepelyansky-08,flach:024101}.
Without nonlinearity $\beta = 0$, eq.~\eqref{eqn:gdanse} is a standard Anderson model describing a disordered lattice. The Hamilton operator can then be diagonalized,  leading to a system of eigenfunctions~$\Phi_{k,n}$ with energy eigenvalues~$\epsilon_k$.
An arbitrary wave function $\psi$ can be decomposed into these eigenfunctions $\psi_n = \sum_k C_k \Phi_{k,n}$, with $C_k(t)=C_k(0)\e^{-i\epsilon_k t}$.
With nonvanishing nonlinearity, this decomposition into eigenfunctions of the linear part of the Hamilton operator is still possible,
but now the coefficients $C_k$ are coupled through a nonlinearity:
\begin{equation} \label{eqn:gdanse_eigen}
 i\frac{\rmd}{\rmd t}C_k = \epsilon_k C_k + \beta \mathcal{N}(C)\;.
\end{equation}
The expression for the nonlinear term $\mathcal{N}(C)$ is rather cumbersome (see~\cite{Mulansky-Diplom} for details).
In the case of integer indices $\alpha = 1,2,3,\dots$ the nonlinear coupling~$\mathcal{N}(C)$ can be simplyfied by introducing an overlap matrix~$V$ of $2\alpha+2$ eigenfunctions.
Using this, eq.~\eqref{eqn:gdanse_eigen} reads as:
\begin{equation*}
\begin{aligned}
 i\dot C_k &= \epsilon_k C_k\\& +\beta\sum_{
\substack{
\widehat n_1 \dots \widehat n_\alpha \\
\widetilde n_1 \dots \widetilde n_\alpha \\
\bar n }} 
 V_{
\substack{
\widehat n_1 \dots \widehat n_\alpha \\
\widetilde n_1 \dots \widetilde n_\alpha \\
\bar n,k }} C_{\widehat n_1} \cdots C_{\widehat n_\alpha} C_{\widetilde n_1}^* \cdots C_{\widetilde n_\alpha}^* C_{\bar n}.
\end{aligned}
\end{equation*}

For the gDANSE model~\eqref{eqn:gdanse} one is interested in the spreading of initially bounded perturbations (\textit{i.e.} at time $t=0$ only several sites of the lattice are excited $\psi_n\neq 0$ while $\psi_n=0$ outside). Because linear modes $\Phi_{k,n}$ are exponentially localized, these initial states are equivalent to an initial seeding of a finite number of modes.
A qualitative picture of the spreading, based on eq.~\eqref{eqn:gdanse_eigen}, looks as follows: the nonlinear coupling leads to chaotic dynamics of excited modes, as result a chaotic force acts on nonexcited modes and leads to their growth, and so on. 
Although several attempts to explain the detailed mechanisms of spreading have been made based on nonlinear eigenmode interactions (see, \textit{e.g.}, \cite{flach:024101} and \cite{PhysRevLett.70.1787}) a convincing, detailed description could not be found, yet. This relates to a still missing general understanding of statistical properties of weak chaos in high-dimensional Hamiltonian systems (cf. concept of fast Arnold diffusion developed by Chirikov and Vecheslavov~\cite{chirikov:1993}). Nevertheless, simplifying assumptions allowed to develop a  phenomenological picture of a slow, subdiffusive spreading \cite{Pikovsky-Shepelyansky-08,PhysRevLett.70.1787,flach:024101}. One of the aims of our work is to test these pictures by numerics.

As the spreading is induced by the nonlinear term in eq.~\eqref{eqn:gdanse}, we suggest as a phenomelogical description the \textit{nonlinear diffusion equation} for the probability density $\rho = |\psi|^2$:
\begin{equation} \label{eqn:nl_diff_eq}
 \frac{\partial \rho}{\partial t} = D \frac\partial{\partial x} \left(\rho^a \frac{\partial \rho}{\partial x}\right),
\end{equation}
where the effective diffusion coefficient obeys a power law dependence $\sim \rho^a$.
Note, that at the moment we do not see a way to derive
\eqref{eqn:nl_diff_eq} from the DNLS \eqref{eqn:gdanse} directly, as for this a detailed theory of the microscopic chaos is needed. Neither do we 
claim that a solution of \eqref{eqn:gdanse} satisfies \eqref{eqn:nl_diff_eq}.
Our hope, however, is that the nonlinear diffusion equation provides a reasonable framework for an average spreading behavior of localized states in the DANSE model, to be compared with numerical findings.

\begin{figure}[t]
   \centering
   \includegraphics[height=0.42\textwidth,angle=270]{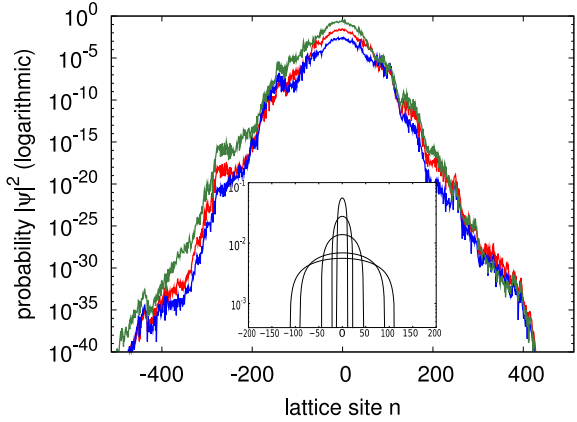}
\caption{(color online) Absolute square of wavefunctions $|\psi_n|^2$ for different nonlinearity indices and times. From top to bottom: $\alpha = 0.5$, $t=10^4$ (green); $\alpha=1$, $t=10^5$ (red); $\alpha = 2$, $t=10^8$ (blue). $W=4.0,\, \beta=1.0,\, N=1024$ in all three cases.
The curves are shifted vertically for a better visibility.
The inset shows the self-similar solution \eqref{eqn:self-sim_solution} of the nonlinear diffusion equation for increasing times $t$ (top to bottom).
}
\label{fig:wf_comp_times}
\end{figure}

Asymptotically, for $a>0$, the spreading in \eqref{eqn:nl_diff_eq} is described by the self-similar solution (see inset in fig.~\ref{fig:wf_comp_times})
\begin{equation}\label{eqn:self-sim_solution}
\rho = 
\begin{cases}
  (Dt)^{-1/(2+a)} \left( A - \frac{ax^2}{2(a+2)t^{2/(2+a)}} \right )^{1/a} &  x<x_0,\\
	0 &  x>x_0,
\end{cases}
\end{equation} 
here $A$ is a constant fixed by the normalization condition $\int \rho\, \rmd x = 1$.
The position of the edge of the spreading field $x_0$  has the following time dependence:
\begin{equation} \label{eqn:diff_pwr_law}
  x_0 = \sqrt{2A\frac{2+a}a(Dt)^{2/(2+a)}}\sim t^{1/(2+a)}.
\end{equation}

In order to characterize the spreading in gDANSE quantitatively and to compare it with the nonlinear diffusion model, one interpretes $\rho_n=|\psi_n|^2$ as probability at site $n$ and uses typically the mean squared deviation
$(\Delta n)^2 = \langle (n-\langle n\rangle)^2\rangle$ and the so-called participation number $P^{-1}=\sum_n \rho_n^2$. Here we suggest to also use \emph{R\'enyi-Entropies} \cite{Renyi:1961} as a new characterization tool:
\begin{equation}
  I_q = \frac1{1-q}\ln \sum_n \rho_n^{q}=\frac1{1-q}\ln \sum_n |\psi_n|^{2q}\;.
\end{equation}
Obviously, $I_{q\rightarrow1} = S=-\sum_n\rho_n \ln \rho_n$ is the usuual Shannon entropy, while the participation number is $P=e^{I_2}$.

For the self-similar evolution governed by the nonlinear diffusion equation~\eqref{eqn:self-sim_solution},
the variance $(\Delta x)^2 = \int x^2 \rho(x) \rmd x$ as well as the R\'enyi entropies $I_q = \frac{1}{1-q}\ln \int \rho(x)^q \rmd x$ can be evaluated analytically:
\begin{align}
 (\Delta x)^2 &=(Dt)^{\frac{2}{2+a}} \frac{2 (2+a) \Beta\left(\frac{a+1}{a},\frac{3}{2}\right)}{a \Beta\left(\frac{a+1}{a},\frac{1}{2}\right)}\;,\\
	I_q &=\frac{1}{2+a}\left(\ln Dt + \ln \left(\frac{2(a+2)}{a}\right)\right)\nonumber\\&-
	\frac{a+2q}{(a+2)(1-q)}
	\ln \Beta\left(\frac{a+1}{a},\frac{1}{2}\right)\nonumber\\&+
	\ln \Beta\left(\frac{a+q}{a},\frac{1}{2}\right)\;,
	\label{eq:iq}
\end{align} 
where $\Beta(x,y)$ is the Beta function. In this self-similar situation all
the entropies grow with the same rate.
Correspondingly, the asymptotic growth indices of the entropies and of the mean square displacement defined as
\begin{equation} \label{eqn:diff_exponents}
  \exp(I_q)\sim t^{\nu_q},\qquad (\Delta x) \sim t^{\nu_{\text{var}}}
\end{equation}
have the same value $\nu_q=\nu_{\text{var}}=\frac{1}{2+a}$.

\vspace{0.2em}
The main goal of introducing R\'enyi entropies as characterization of the spreading is to control the peak structure of the field, in a similar way as these entropies are used in the multifractal formalism.
Indeed, the parameter $q$ determines the sensitivity of $I_q$ on the peaks of the distribution $\rho_n$.
Larger values of $q$ emphasize the high peaks while for small values of $q$ the background of the distribution governs $I_q$ (the entropy $I_0$ characterizes the support of the distribution). Therefore, if there are large peaks that do not spread (but, \textit{e.g.}, just drift), then the R\'enyi entropies with large $q$ will not grow. Following the evolution of these entropies, we can visualize changes in the peak structure of the distribution.

Fig.~\ref{fig:wf_comp_times} shows exemplary averaged wave functions for three different values of $\alpha=1,2,3$ at times $t=10^4$, $10^5$ and $10^8$.
One clearly still sees the peaked plateau even though these wave functions were already averaged over time windows and disorder realizations.

\begin{figure}[t]
   \centering
   \psfrag{xlabel}{$\log_{10} t$}
   \psfrag{ylabel}{entropies $I_q$}
   \includegraphics[height=0.3\textwidth]{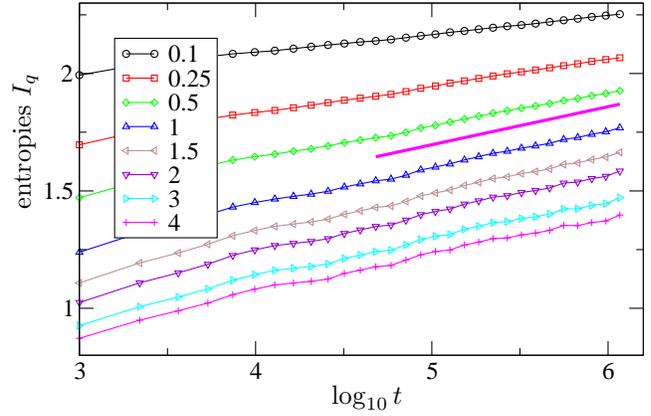}
\caption{(color online) Evolution of the entropies $I_q$ (calculated using logarithm with base 10) in the gDANSE \eqref{eqn:gdanse} with $\alpha=\beta=1$.
The solid line has slope $0.162$, it is drawn as a linear fit for the growth of the Shannon entropy $I_1=S$.
}
\label{fig:entropies}
\end{figure}

In fig.~\ref{fig:entropies} we show the evolution of R\'enyi entropies for the ``standard'' nonlinearity index $\alpha=1$. We stress that for these calculations no averaging of the wave function was performed, instead instantaneous entropies have been averaged over time (over time intervals between succesive markers in the plot) and realizations of disorder. All entropies with $q\geq 0.5$ show almost the same growth rate for large times. Entropies with very small indices $q=0.1,\;0.25$ grow slightly slower, but this is not really relevant: these entropies effectively measure the support of the distribution and are dominated by highly fluctuating exponentially decaying tails of localized eigenmodes  $\Phi_{k,n}$.

Another way to characterize the peak structure of the  distributions
is to look at differences between  R\'enyi entropies. The mostly suitable choice appears to be the structural entropy $S_\text{str}$ introduced together with the localization entropy $S_\text{loc}$ in  \cite{PhysRevA.46.3148}:
\begin{align}
 S_\text{str} &= I_1-I_2=S - \ln P\\
 S_\text{loc} &= I_2 = \ln P.
\end{align}
From this definition, we can say that the entropy is build from a localization part and a structural part ${S = S_\text{loc} + S_\text{str}}$.
To see the meaning of the structural entropy, we calculate it for a uniform distribution of length $L$. In this case  we find ${S=\ln P=\ln L}$ and $S_\text{str}=0$. The (always positive) values of the structural entropy measure the relative nonuniformity of a distribution.
For the self-similar solution~\eqref{eqn:self-sim_solution} of the nonlinear diffusion equation the structural entropy, according to \eqref{eq:iq}, is constant.

\begin{figure}[t]
   \centering
   \includegraphics[height=0.42\textwidth,angle=270]{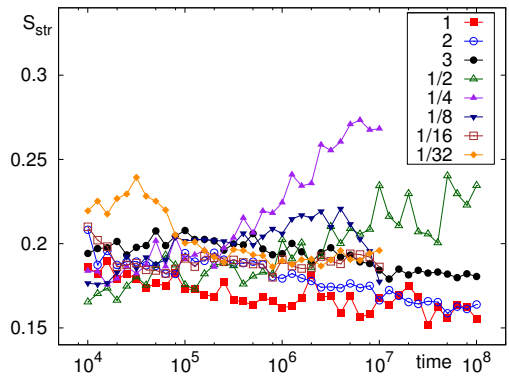}
\caption{(color online) Structural entropy $S_\text{str}$ vs.\ time for different values of $\alpha$ (see label).
Parameters values are ${W=4},\, {\beta=1.0},\, {N=1024}$.
}
\label{fig:s_str_time}
\end{figure}

\begin{figure}[t]
   \centering
   \includegraphics[height=0.45\textwidth,angle=270]{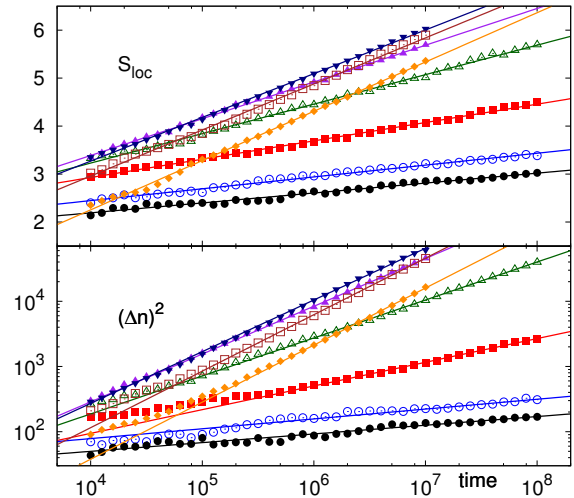}
\caption{(color online) Time evolution of the localization entropy~$S_\text{loc}$ (top panel) and the the second moment~$(\Delta n)^2$ (bottom panel) in the gDANSE model for different values of the nonlinearity index $\alpha$ (see labels in fig.~\ref{fig:s_str_time}).
Other parameters were $W=4,\, \beta=1.0,\, N=1024$.
The lines are numerical fits (at the final stage of the time evolution) $S_\text{loc} \sim \nu_2\ln t$ and $(\Delta n)^2 \sim t^{2\nu_\text{var}}$ respectively. Fitting results are plotted in fig.~\ref{fig:nu_alpha}.}
\label{fig:dn_time_all}
\end{figure}

Fig.~\ref{fig:s_str_time} shows the time dependence of the structural entropy for different values of the nonlinearity index $\alpha$ obtained from numerical integration of \eqref{eqn:gdanse} for an initially localized wavefunction (details on the numerical integration scheme follow later in this text).
One can see that $S_\text{str}$ remained rather constant over time while the localization entropy increased as $\ln P \sim \nu_2 \ln t$ (fig.~\ref{fig:dn_time_all}).
However, $\alpha = 1/2$ and $\alpha = 1/4$ showed a stronger increase of $S_\text{str}$ than the other values of $\alpha$.
But compared to $S_\text{loc}$, which exhibits a clear, straight growth over time, the increase of the structural entropy is rather small.
These findings demonstrate that the peak structure of the spreading wave function remains constant with time, while the entropies grow as expected from the delocalization effect induced by the nonlinearity. The wave packet spreads uniformly and is not dominated by a few peaks. This supports  validity of the nonlinear diffusion equation as a suitable model for the spreading in gDANSE.

Given the nonlinear diffusion equation as a suitable phenomenological description of the long-time behavior of the gDANSE model, one still has to find the relation between the nonlinear term in the gDANSE~\eqref{eqn:gdanse} and the one in the diffusion equation~\eqref{eqn:nl_diff_eq}.
More precisely, the relation of $\alpha$ and $a$ is unknown.
In the literature, three approaches have been discussed:\\
\begin{description}
  \item[A) $a=2\alpha$] which will be called \emph{strong stochasticity} here and the resulting spreading exponent $\nu_A = 0.5/(1+\alpha)$ was derived
by Flach \textit{et al.}~\cite{flach:024101} under the assumption of completely random phases of the eigenmode amplitudes $C_k$.
  \item[B) $a=3\alpha$] (\emph{weak stochasticity}) gives the spreading exponent as $\nu_B=1/(2+3\alpha)$, this law corresponds to early results of Shepelyansky \cite{PhysRevLett.70.1787} for the standard DANSE model and for the quantum kicked rotor model with nonlinearity.
  \item[C) $a=4\alpha$] (\emph{very weak stochasticity}) leads to the spreading exponent $\nu_C = 0.5/(1+2\alpha)$, this result was obtained by Flach \textit{et al.}~\cite{flach:024101} by applying some arguments on reduced chaoticity of the excited modes compared to Shepelyansky's model.
 \end{description}
Note, that in the ``standard'' case $\alpha=1$ these models yield $\nu_A=1/4$,
$\nu_B=1/5$ and $\nu_C=1/6$, respectively.

To find the spreading exponents we have performed 
extensive numerical simulations for different nonlinearity indices $\alpha = 1/32,\, 1/16,\, 1/8,\, 1/4,\, 1/2,\, 1,\, 2,\, 3$.
We took 10 disorder realizations with disorder strenght $W=4$ and lattice size $N=1024$ and initialized them with a single excitation at one lattice site $\psi_{n_0} = 1$.
It was found recently that the energy of the state, which is a conserved quantity, is crucial for the spreading behavior \cite{mulansky:2009}.
Therefore, we ensured the energy of the states to be $|E| < 1$ for all of the initial conditions by artificially setting the potential value to zero at the starting point $V_{n_0} = 0$.
Hence, we are always in the center of the energy band where no breather should interfere with the spreading behavior.
Then we ran the numerical time evolution based on an operator splitting and the Crank-Nicolson scheme for the linear part.
For the time discretization we used a step size of $\Delta t = 0.1$.
$\beta$ was set to $1.0$ throughout all simulations.
This integration method is unitary and hence preserves the total probability within the computer accuracy $10^{-16}$ and the energy was fluctuating less than 1\% during the simulations.
These simulations were done for each disorder realization and we computed $P$, $(\Delta n)^2$ and $S_\text{str}$ for times between $t=10^4$ to $t=10^8$ ($10^7$ for $\alpha<1$) and averaged over (exponentially growing) time windows.
Finally, we fitted $(\Delta n)^2 \sim t^{2\nu_{\text{var}}}$ and $P\sim t^{\nu_2}$ for each realization separately and then averaged the results over disorder realizations.
This was repeated for each value of the nonlinearity index ${\alpha}$.

\begin{figure}[t]
   \centering
   \includegraphics[height=0.42\textwidth,angle=270]{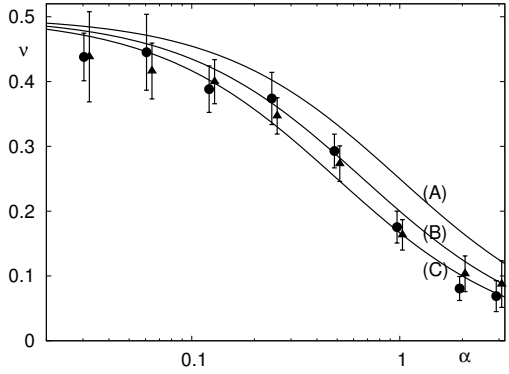}
\caption{Exponents of the spreading law in dependence of the nonlinearity index $\alpha$.
Circles are the values obtained from the numerical fits of $(\Delta n)^2 \sim t^{2\nu_\text{var}}$ and the triangles come from the fits of $S_\text{loc} \sim \nu_2\ln t$ (compare fig.~\ref{fig:dn_time_all}).
The values are slightly shifted horizontally for a better distinguishability.
The solid lines are the exponents given by \eqref{eqn:diff_exponents} for the three stochasticity assumptions:
A)~\emph{strong stochasticity}, B)~\emph{weak stochasticity}, C)~\emph{very weak stochasticity}.}
\label{fig:nu_alpha}
\end{figure}

In fig.~\ref{fig:dn_time_all}, the results of these simulations are shown.
For all values of $\alpha$ subdiffusive spreading has been found, allowing us to obtain $\nu_{\text{var}}$ and $\nu_2$ for different values of~$\alpha$.
In fig.~\ref{fig:nu_alpha} the numerical results for the spreading exponent are compared with the exponents derived from the nonlinear diffusion equation for the different assumptions A), B) and C).
The fits for $\nu_{\text{var}}$  (points) and $\nu_2$ (triangles) gave very similar results and are both close to the theoretical estimates $\nu_B$ and $\nu_C$.
The values $\nu_A$ are clearly larger than the numerical results for all values of $\alpha$.
Note, that the spreading for $\alpha=2,3$ is quite slow and so the numerical fits can not give very relieable results for these parameter values.

\begin{figure}[t]
   \centering
   \includegraphics[height=0.24\textwidth,angle=270]{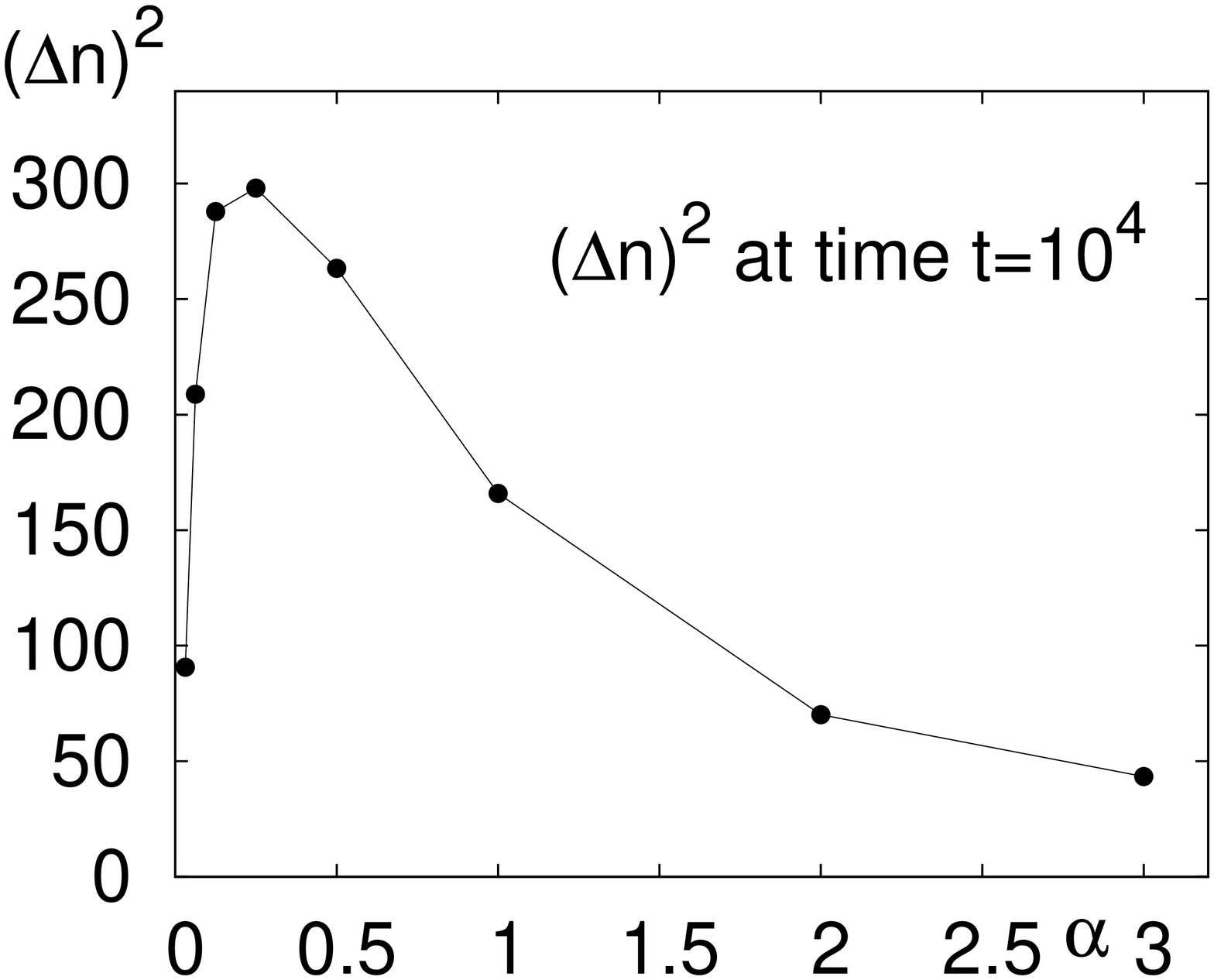}
   \includegraphics[height=0.24\textwidth,angle=270]{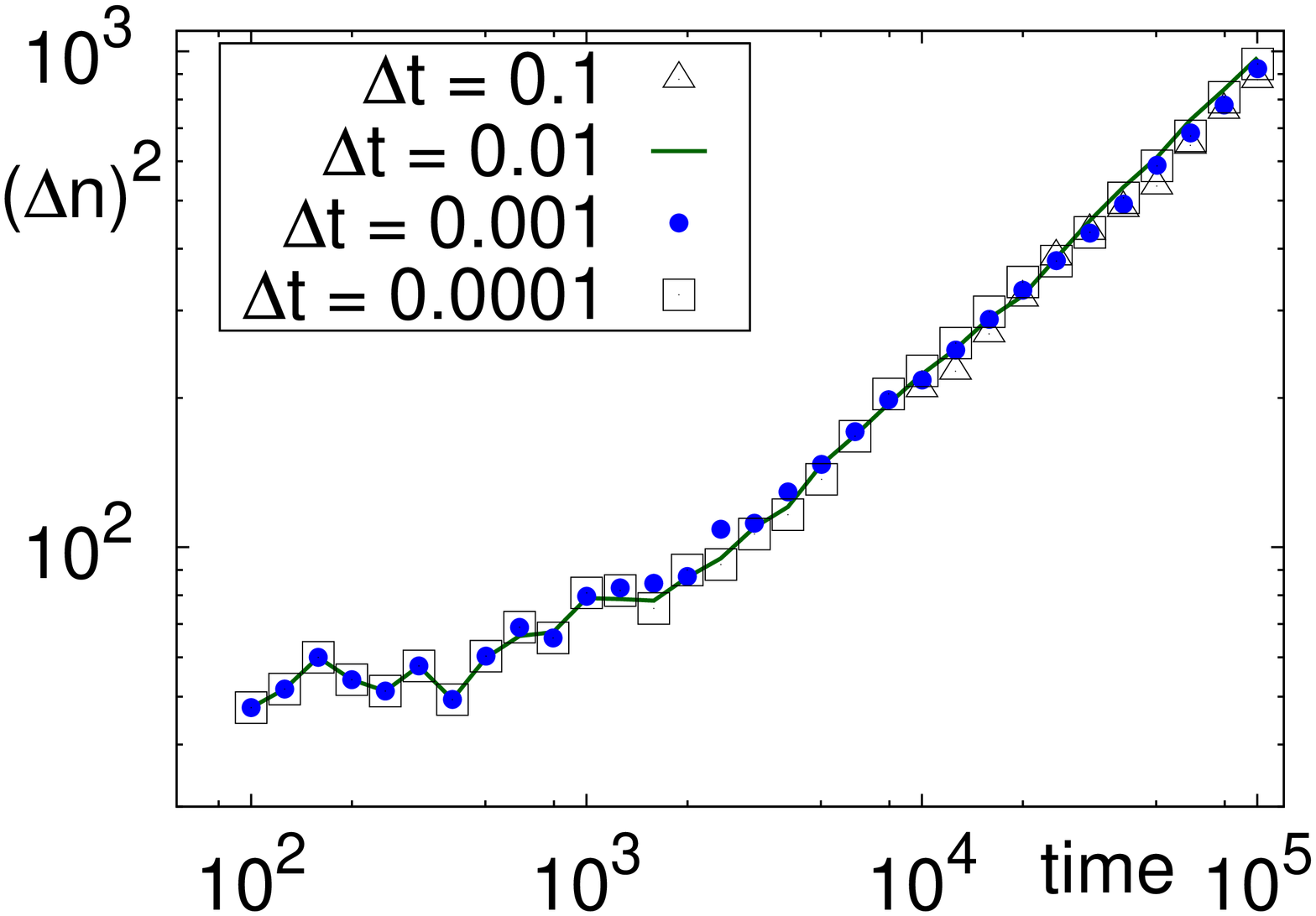}
\caption{Left Panel: Second moment~$(\Delta n)^2$ at time~$t=10^4$ vs.\ nonlinearity index~$\alpha$.
Note the non-logarithmic scaling of the $\alpha$-axis for a better comparability with~\cite{alpha_b_p:2009}. Right Panel: Initial time evolution of the second moment~$(\Delta n)^2$ for $\alpha = 1/16$ and different values of time discretization $\Delta t$.
}
\label{fig:dn_t1E4_dt}
\end{figure}

As $\alpha$ approaches zero, all estimates A)--C) converge to one $\nu_{A,B,C}(\alpha\rightarrow 0) \rightarrow 1$, as the nonlinear diffusion equation becomes the usual linear one in this limit.
For the original model, in contrast, the nonlinear term changes to a linear one when ${\alpha \rightarrow 0}$ and no spreading at all should be observed as the gDANSE becomes a linear equation with the Anderson localization property. The riddle is resolved by noting the role of the constant $D(\alpha)$ in the nonlinear diffusion equation~\eqref{eqn:nl_diff_eq}. According to the Anderson localization picture, we have to set $D(0)=0$, what assures no spreading in the linear case.
This picture corresponds to the recent results by Veksler \textit{et al.}~\cite{alpha_b_p:2009}. They found decreasing spreading exponents for $\alpha \rightarrow 0$ when investigating the short time behavior up to $t = 10^4$ \cite{alpha_b_p:2009}.
A closer look on fig.~\ref{fig:dn_time_all} also reveals that for very small $\alpha = 1/16, 1/32$ the spreading seems to speed up between $t=10^4$ and $t=10^5$.
The graph in the right panel of fig.~\ref{fig:dn_t1E4_dt} clearifies this as one sees that for $\alpha=1/16$ the spreading is delayed roughly up to $10^4$.
Furthermore, this plot examplarily shows the independence of our results on the time discretization $\Delta t$.
Additionally, we have plotted the values of the second moment at time $t=10^4$ (left panel in fig.~\ref{fig:dn_t1E4_dt}) for the different nonlinearity indices~$\alpha$ (note non-logarithmic scaling of $\alpha$ for a better comparability with \cite{alpha_b_p:2009}) and the decreasing of $(\Delta n)^2$ supports the conclusion that $D\to 0$ for small $\alpha$.
One clearly sees the maximum at $\alpha = 0.25$, which corresponds to the maximum of the initial spreading exponent found in \cite{alpha_b_p:2009}.
Our hypothesis is that the behavior of $D(\alpha)$ could be estimated via a calculation of Lyapunov exponents of chaos, to be reported elsewhere.

The main conclusion of this paper is that the  spreading of initially localized states in nonlinear disordered lattices can be phenomenologically well described by self similar solutions of the nonlinear diffusion equation.
Its validity is supported by the finding that different R\'enyi entropies grow with the same exponent. In particular, the structural entropy~$S_\text{str}$ was used to measure the relation of the peaks and the background field in the spreading states.
This quantity was found to remain rather constant during the spreading in most of the cases, supporting thus the self-similarity in average of the, however strongly fluctuating and highly peaked, wave function.
We have shown  the nonlinear diffusion equation to be applicable also in the linear limit of vanishing nonlinearity index if one assumes that the diffusion coefficient vanishes in this limit as well. 
We have found numerically that models of \emph{weak} and \emph{very weak stochasticity} give good approximations for the spreading exponent~$\nu$, but based on our numerics we cannot discriminate them (cf.~\cite{flach:2010}). Here, additional studies of  microscopic statistical properties of the underlying chaos are needed. Quite recently, it was suggested that a crossover between strong and very weak chaos may occur in the DANSE model~\cite{flach:2010}. In our simulations, we could not identify such a crossover.
In a future work a much larger range of system parameters should be explored in a search for such an effect.

We acknowledge useful discussions with K.~Ahnert, M.~Abel, D.~Shepelyansky, S.~Flach, and S.~Fishman.

\end{document}